\documentclass[preprint,preprintnumbers,amsmath,amssymb,
superscriptaddress,floatfix]{revtex4}
\usepackage{graphicx}
\usepackage{amsmath}
\usepackage{bm}
\usepackage{mathrsfs}
\usepackage{color}
\usepackage{slashed}
\usepackage{dcolumn}
\usepackage[normalem]{ulem}

\newcommand{\be}{\begin{equation}}
\newcommand{\ee}{\end{equation}}
\newcommand{\ba}{\begin{eqnarray}}
\newcommand{\ea}{\end{eqnarray}}

\begin{document}

\title{The $\alpha$ condensate states of atomic nuclei ${}^{12}C$, ${}^{16}O$ and ${}^{20}Ne$ in an analytical solvable model}

\author{Bao-Xi Sun}
\email{sunbx@bjut.edu.cn}
\affiliation{School of Physics and Optoelectronic Engineering, Beijing University of Technology, Beijing 100124, China}

%\date{\today}

\begin{abstract}
The $\alpha$ condensation in the ${}^{12}C$, ${}^{16}O$ and ${}^{20}Ne$ nuclei is investigated within an analytical solvable model.
It is found that the calculated ratio of the ground state energies of the Hoyle state of ${}^{12}C$ and the Hoyle-like state of ${}^{16}O$ is consistent with that of the experimental values. Along this clue, the ground state energy of ${}^{20}Ne$ is obtained  to be 1MeV approximately, which is far less than the experimental value of 3MeV. Additionally, the root-mean-square radii of these nuclei are also calculated, and all of them lies around 9fm, which is different from the result calculated with the Tohsaki-Horiuchi-Schuck-Ropke(THSR) wave function. Since the root-mean-square radius is relevant to the ground state energy of the $\alpha$ condensate nucleus, the root-mean-square radii of ${}^{16}O$ and ${}^{20}Ne$ are also calculated with the ground state energies used in the THRS wave function. As a result, the root-mean-square radii of ${}^{16}O$ and ${}^{20}Ne$ reduced to 5fm, and it is similar to the result obtained with the THRS wave function. The calculation result manifests that the root-mean-square radius of $\alpha$ condensate nuclei decreases with the energy increasing.
\end{abstract}


\maketitle

\section{Introduction}
\label{sect:intoduction}

The Hoyle state of ${}^{12}C$ plays a central role in the stellar nucleosynthesis, which is composed of three $\alpha$ particles and  7.65MeV higher than the ground energy of ${}^{12}C$\cite{Hoyle1954}.
Since the $\alpha$ particle(${}^{4}He$) consisting of two neutrons and two protons is stable sufficiently and assumed to be a building block in light nuclei, especially in the 4$n$ nuclei, such as ${}^{16}O$ and ${}^{20}Ne$.
The $\alpha$ particles are composite bosons and the interaction of them is weak, therefore, it is more possible that all $\alpha$ particles fall to the lowest energy state, and the $\alpha$ condensation comes into being in light nuclei, just as the Bose-Einstein condensation of bosonic atoms in magneto-optical traps\cite{Schuck2008-4}.
Along this clue, the $\alpha$ condensate states of ${}^{12}C$ and ${}^{16}O$ are proposed by means of the Tohsaki-Horiuchi-Schuck-Ropke(THSR) wave function\cite{Schuck2001,Schuck2008}.
Reasonably, the Hoyle state of ${}^{12}C$ is confirmed as a $3\alpha$ condensate state, and corresponds to the second $J^P=0^+$ state in the ${}^{12}C$ nucleus, which is located at $0.38$MeV above the 3$\alpha$ threshold energy\cite{Schuck2001}.
As for the ${}^{16}O$ nucleus,
altogether six $J^P=0^+$ state of ${}^{16}O$ with different energies up to 15.1MeV have been reported experimentally\cite{Schuck2008-34,Schuck2008-35}, and all of them are studied in the orthogonality condition model with the THSR wave function\cite{Schuck2008}.
It is concluded that the second $J^P=0^+$ state at 6.05MeV and the third $J^P=0^+$ state at 12.1MeV have $\alpha+{}^{12}C$ structures  where the $\alpha$ particle orbits around the ${}^{12}C(0^+_1)$ core in a $S$ wave and around ${}^{12}C(2^+_1)$ core in a $D$ wave, respectively.
The sixth $J^P=0^+$ state of ${}^{16}O$ is assumed to be a $4\alpha$ condensate state with a large root-mean-square radius about 5fm. However, the calculated energy of the sixth $J^P=0^+$ state of ${}^{16}O$ is about 2MeV higher than the $4\alpha$ threshold, and the result is different from the experimental energy of the sixth $J^P=0^+$ state of at 15.1MeV, which is only 0.66MeV above the $4\alpha$ threshold\cite{ZhouBo35}.

In the case of ${}^{20}Ne$, three states at $E_x=23.6$MeV, 21.8MeV and 21.2 MeV are observed in the experiment of inelastic ${}^{20}Ne(\alpha,\alpha^\prime){}^{20}Ne$ reaction\cite{ZhouBo35}, which couple strongly to the sixth $J^P=0^+$ state in ${}^{16}O$. Meanwhile, in the reaction ${}^{22}Ne(p,t){}^{20}Ne$. the excited states up to 25MeV of ${}^{20}Ne$ have been studied, and a state at $E_x=22.5$MeV can not be interpreted in the nuclear shell model and might be a 5$\alpha$ cluster state\cite{ZhouBo36}.
Correspondingly, the $5\alpha$ condensation in ${}^{20}Ne$ is also investigated with the THSR wave function,
and two $J^P=0^+$ states of ${}^{20}Ne$ are obtained theoretically, which are 3MeV above the 5$\alpha$ threshold at 19.2MeV and assumed to be $5\alpha$ condensate states\cite{ZhouBo}.

The three-body problem has been evaluated rigorously by variational calculations to study the spectrum of exotic baryons\cite{Richard90-1,Richard90-2,Richard93,Basdevant96}.
Finally, this approach is summarized in a problem set of quantum mechanics\cite{Book}. When a harmonic oscillator potential between two particle is taken into account, the three-body wave function can be obtained exactly in the Jacobi variables. In this case, the three-body ground state energy equals to three times of the two-body energy with a reduced mass, which is three fourth of the original particle mass exactly.

Apparently, it is reasonable to apply this approach to study the structure of Hoyle state of ${}^{12}C$. Furthermore,
this method is extended to solve the four-body and five-body problems, and the $\alpha$ condensate states of ${}^{16}O$ and ${}^{20}Ne$ are also investigated respectively. The corresponding ground state energies and root-mean-square radii are calculated precisely.

The whole article is organized as follows. The theoretical framework to study three-particle, four-particle and five-particle systems are evaluated respectively in Section~\ref{Framework}, and the ground state energies and root-mean-square radii of ${}^{12}C$, ${}^{16}O$ and ${}^{20}Ne$ are calculated in Section~\ref{sect:results}. Finally, a summary is given in Section~\ref{sect:summary}.

\section{Framework}
\label{Framework}

\subsection{The three-particle system with a harmonic oscillator potential}
\label{sect:3body}

In a three-particle system where the particle masses are the same as each other, the total Hamiltonian of the three-particle system is
\be
\label{eq:202412031456}
\hat{H}^{(3)}=\frac{\vec{p}_1^2}{2m}+\frac{\vec{p}_2^2}{2m}+\frac{\vec{p}_3^2}{2m}+V(r_{12})+V(r_{13})+V(r_{23}),
\ee
where $\vec{p}_1$, $\vec{p}_2$ and $\vec{p}_3$ are the three-momenta of the particles, and $r_{12}=|\vec{r}_1-\vec{r}_2|$, $r_{13}=|\vec{r}_1-\vec{r}_3|$ and $r_{23}=|\vec{r}_2-\vec{r}_3|$ are distances of them, respectively.
Since three particles have the same mass, the Hamiltonian in Eq.~(\ref{eq:202412031456}) can be rewritten as
\be
\label{eq:202412031513}
\hat{H}^{(3)}=\frac{\vec{P}^2}{2M}+\hat{H}_{12}+\hat{H}_{13}+\hat{H}_{23},
\ee
with
\be
\hat{H}_{ij}=\frac{(\vec{p}_i-\vec{p}_j)^2}{6m}+V(r_{ij})=\frac{[(\vec{p}_i-\vec{p}_j)/2]^2}{2\mu}+V(r_{ij}),
\ee
and the reduced mass $\mu=3m/4$, the total momentum of the system $\vec{P}=\vec{p}_1+\vec{p}_2+\vec{p}_3$, and the total mass $M=3m$.

It is apparent that the first term in Eq.~(\ref{eq:202412031513}) corresponds to the kinetic energy of the center of mass of the three-particle system, which is not relevant to the inner structure of the system. Moreover, the two-body Hamiltonians $\hat{H}_{12}$, $\hat{H}_{13}$ and $\hat{H}_{23}$ do not commute with each other, therefore, the energy of the three-particle system could not be the summation of energies of three two-particle systems.

In order to obtain the eigen-energy of the Hamiltonian of the three-particle system, an orthogonal transformation must be performed to positions and momenta of three particles. Supposing $r^T=\{ \vec{r}_1, \vec{r}_2, \vec{r}_3  \}$ and $p^T=\{ \vec{p}_1, \vec{p}_2, \vec{p}_3  \}$, the Jacobi variables are introduced as $R=Ar$ and $Q=Ap$, with $R^T=\{ \vec{R}_1, \vec{R}_2,  \vec{R}_3  \}$, $Q^T=\{ \vec{Q}_1, \vec{Q}_2, \vec{Q}_3  \}$, the orthogonal matrix
\be
A=
\left( \begin{array}{ccc}
 \frac{1}{\sqrt{2}} & -\frac{1}{\sqrt{2}} & 0  \\
-\frac{1}{\sqrt{6}} & -\frac{1}{\sqrt{6}} & \frac{2}{\sqrt{6}} \\
 \frac{1}{\sqrt{3}} &  \frac{1}{\sqrt{3}} & \frac{1}{\sqrt{3}} \\
\end{array}\right),
\ee
and $A^T A=I$.
According to the communication relations $[r_j^\alpha, p_k^\beta]=i \hbar \delta_{jk} \delta_{\alpha \beta}$ with $\alpha, \beta=1,2,3$, it is easy to verify that Jacobi variables satisfy canonical communication relations:
\be
\left[\hat{R}^\alpha_j, \hat{Q}^\beta_k \right]=i \hbar \delta_{jk} \delta_{\alpha \beta}.
\ee

With the orthogonal transformation, it is easy to obtain
\be
\vec{Q}_1^2+\vec{Q}_2^2+\vec{Q}_3^2=\vec{p}_1^2+\vec{p}_2^2+\vec{p}_3^2,
\ee
\be
\label{eq:202501061112}
\vec{R}_1^2+\vec{R}_2^2+\vec{R}_3^2=\vec{r}_1^2+\vec{r}_2^2+\vec{r}_3^2,
\ee
and
\be
\label{eq:202501061202}
3\left( \vec{R}_1^2+\vec{R}_2^2 \right)=(\vec{r}_1-\vec{r}_2)^2+(\vec{r}_1-\vec{r}_3)^2+(\vec{r}_2-\vec{r}_3)^2.
\ee

If a harmonic oscillator potential $V(r)=\frac{1}{2} \kappa r_{ij}^2$ between two particles is assumed, the Hamiltonian of the three-particle system can be written in the form of Jacobi variables,
\be
\hat{H}^{(3)}=\hat{H}_1+\hat{H}_2+\hat{H}_{cm},
\ee
with $\hat{H}_{cm}=\frac{Q_3^2}{2m}=\frac{P^2}{6m}$ the center of mass Hamiltonian and
\be
\hat{H}_i=\frac{Q^2_i}{2m}+\frac{3}{2} \kappa R^2_i,~~~~i=1,2.
\ee
Apparently, $\hat{H}_1$, $\hat{H}_2$ and $\hat{H}_{cm}$ commute with each other, and the eigenvalue and eigenfunction of the three-particle Hamiltonian $\hat{H}^{(3)}$ can be obtained easily.
Therefore, the ground state energy of the three-particle system in the center of mass frame takes form of
\be
\label{eq:202412041008}
E^{(3)}=3 \sqrt{3} \hbar \sqrt{\frac{\kappa}{m}}.
\ee
Moreover, the ground state energy of the three-particle system $E^{(3)}$ could be written as three times of the ground state energy of a two-particle system with a reduced mass $\mu=\frac{3}{4}m$, i.e.,
\be
E^{(3)}=3 E^{(2)}(\mu),
\ee
with $E^{(2)}(\mu)=\frac{3}{2} \hbar \sqrt{\frac{\kappa}{\mu}}$.

When the orthogonal transformation is performed, the ground state wave function of ${}^{12}C$ can be written as
\be
\psi(\vec{R}_1,\vec{R}_2)=\frac{1}{4\pi} \mathcal{R}(R_1)\mathcal{R}(R_2),
\ee
where the radial part of the wave function is
\be
\label{eq:202501061113}
\mathcal{R}(R_1)=\alpha^{3/2} \left(\frac{4}{\sqrt{\pi}} \right)^{1/2} \exp{-\frac{\alpha^2 R_1^2}{2}},
\ee
with
\be
\label{eq:01061318}
\alpha=\left(\frac{\sqrt{3\kappa m}}{\hbar} \right)^{1/2}.
\ee
In the center of mass frame, $\vec{R}_3=0$, therefore the root-mean-square radius of ${}^{12}C$ can be calculated with the radial wave function in Eq.~(\ref{eq:202501061113}) and the relation in Eq.~(\ref{eq:202501061112}), i.e.,
\be
\label{eq:202501061210}
\sqrt{\langle \vec{r}^2 \rangle}=\left(\frac{1}{3} \sum_{i=1}^3 \langle{ \vec{r}_i^2 \rangle}\right)^{1/2}=\left(\frac{1}{3} \sum_{i=1}^2 \langle \vec{R}_i^2 \rangle \right)^{1/2}=\sqrt{\frac{2}{3} \langle \vec{R}_1^2 \rangle},
\ee
where $\langle \vec{r}_1^2 \rangle=\langle \vec{r}_2^2 \rangle=\langle \vec{r}_3^2 \rangle$ and $\langle \vec{R}_1^2 \rangle=\langle \vec{R}_2^2 \rangle$ are assumed.

Similarly, the root-mean-square distance of two $\alpha$ particles can be obtained according to Eq.~(\ref{eq:202501061202}),
\be
\label{eq:202501061212}
\sqrt{\langle (\vec{r}_1-\vec{r}_2)^2 \rangle} = \sqrt{2 \langle \vec{R}_1^2 \rangle},
\ee
with $\langle (\vec{r}_1-\vec{r}_2)^2 \rangle=\langle (\vec{r}_1-\vec{r}_3)^2 \rangle=\langle (\vec{r}_2-\vec{r}_3)^2 \rangle$ assumed. Actually, Eq.~(\ref{eq:202501061212}) is also suitable for the four-body system.

With the root-mean-square radius of ${}^{12}C$ in Eq.~(\ref{eq:202501061210}) and the root-mean-square distance of $\alpha$ particles in Eq.~(\ref{eq:202501061212}), the included angle of two $\alpha$ particles in ${}^{12}C$ is
$\theta=120^o$,
which indicates that the three $\alpha$ particles in the Hoyle state of ${}^{12}C$ distribute in a form of an equilateral triangle.

The expectation of $\vec{R}_1^2$ can be calculated with the radial wave function in Eq.~(\ref{eq:202501061113}), i.e.,
\be
\label{eq:01061514}
\langle \vec{R}_1^2 \rangle = \int_0^\infty \mathcal{R}^2(R_1) R_1^4 d R_1=\frac{3}{2\alpha^2}.
\ee
As far as Eqs.~(\ref{eq:202412041008}) and (\ref{eq:01061318}) are concerned, the expectation of $\vec{R}_1^2$ can be written as the function of the ground state energy of ${}^{12}C$, i.e.,
\be
\langle \vec{R}_1^2 \rangle=\frac{9\hbar^2}{2mE^{(3)}},
\ee
 with $m$ the $\alpha$ particle mass.

\subsection{The four-particle system with a harmonic oscillator potential}
\label{sect:4body}

The method in Section~\ref{sect:3body} can be extended to study the four-particle system where the particles have the same mass. In this case, the orthogonal matrix is
\be
A=
\left( \begin{array}{cccc}
 \frac{1}{\sqrt{2}} & -\frac{1}{\sqrt{2}} & 0                  & 0  \\
 \frac{1}{\sqrt{6}} &  \frac{1}{\sqrt{6}} & -\frac{2}{\sqrt{6}}& 0  \\
 \frac{\sqrt{3}}{6} &  \frac{\sqrt{3}}{6} & \frac{\sqrt{3}}{6} & -\frac{\sqrt{3}}{2}  \\
 \frac{1}{2} &  \frac{1}{2} & \frac{1}{2} & \frac{1}{2}  \\
\end{array}\right),
\ee
and similarly,
\be
\vec{Q}_1^2+\vec{Q}_2^2+\vec{Q}_3^2+\vec{Q}_4^2=\vec{p}_1^2+\vec{p}_2^2+\vec{p}_3^2+\vec{p}_4^2,
\ee
\be
\vec{R}_1^2+\vec{R}_2^2+\vec{R}_3^2+\vec{R}_4^2=\vec{r}_1^2+\vec{r}_2^2+\vec{r}_3^2+\vec{r}_4^2,
\ee
and
\be
4\left( \vec{R}_1^2+\vec{R}_2^2+\vec{R}_3^2 \right)=(\vec{r}_1-\vec{r}_2)^2+(\vec{r}_1-\vec{r}_3)^2+(\vec{r}_1-\vec{r}_4)^2
+(\vec{r}_2-\vec{r}_3)^2+(\vec{r}_2-\vec{r}_4)^2+(\vec{r}_3-\vec{r}_4)^2.
\ee
If a harmonic oscillator potential $V(r)=\frac{1}{2} \kappa r_{ij}^2$ between two particles is assumed, the total Hamiltonian of the four-particle system can be written as
\be
\hat{H}^{(4)}=\hat{H}_1+\hat{H}_2+\hat{H}_3+\hat{H}_{cm},
\ee
where the center of mass Hamiltonian $\hat{H}_{cm}=\frac{\vec{Q}_4^2}{2m}$ and $\hat{H}_i=\frac{Q^2_i}{2m}+2 \kappa R^2_i$ with $i=1,2,3$. Apparently, $\hat{H}_{cm}$, $\hat{H}_1$, $\hat{H}_2$, and $\hat{H}_3$ commute with each other.

In the center of mass frame, the ground state energy of the four-particle system takes the form of
\be
\label{202501041347}
E^{(4)}=3E^{(2)}(\mu \rightarrow m, \kappa \rightarrow 4\kappa)=9 \hbar \sqrt{\kappa}{m}.
\ee
If the ground state energy of the four-particle system is treated to be the summation of the energies of four harmonic oscillators, it can be rewritten as
\be
\label{eq:202412051740}
E^{(4)}=%4E^{(2)}(\mu=\frac{4m}{9})=4*\frac{3}{2} \hbar \sqrt{\frac{\kappa}{\mu}}
4E^{(2)}(\mu),
\ee
with the reduced mass $\mu=\frac{4}{9}m$.

The corresponding root-mean-square radius of the four-particle system can be calculated similarly to the case of the three-particle system,
\be
\label{eq:202501061458}
\sqrt{\langle \vec{r}^2 \rangle}=\sqrt{\frac{3}{4} \langle \vec{R}_1^2 \rangle},
\ee
while the root-mean-square distance of two particles still takes the form in Eq.~(\ref{eq:202501061212}). Thus the included angle of two particles is obtained, which is
$\theta=109.47^o$.
Apparently, the four particles form a regular tetrahedron in the coordinate space.

By replacing $\alpha^2=\frac{\sqrt{4\kappa m}}{\hbar}$ into Eq.~(\ref{eq:01061514}), the expectation of $\vec{R}_1^2$ in the four-particle system can be written as
\be
\langle \vec{R}_1^2 \rangle=\frac{27\hbar^2}{4mE^{(4)}},
\ee
with $E^{(4)}$ the ground state energy of the four-particle system in Eq.~(\ref{202501041347}).

\subsection{The five-particle system with a harmonic oscillator potential}
\label{sect:5body}

The five-particle system can be studied with the orthogonal matrix
\be
A=
\left( \begin{array}{ccccc}
 \frac{1}{\sqrt{2}} & -\frac{1}{\sqrt{2}} & 0                  & 0  & 0 \\
 \frac{1}{\sqrt{6}} &  \frac{1}{\sqrt{6}} & -\frac{2}{\sqrt{6}}& 0  & 0 \\
 \frac{1}{2 \sqrt{3}} & \frac{1}{2 \sqrt{3}} & \frac{1}{2 \sqrt{3}} & -\frac{\sqrt{3}}{2} & 0 \\
 \frac{1}{2 \sqrt{5}} &  \frac{1}{2 \sqrt{5}} & \frac{1}{2 \sqrt{5}} & \frac{1}{2 \sqrt{5}} & -\frac{2}{\sqrt{5}} \\
 \frac{1}{\sqrt{5}} &  \frac{1}{\sqrt{5}} & \frac{1}{\sqrt{5}} & \frac{1}{\sqrt{5}} & \frac{1}{\sqrt{5}} \\
\end{array}\right),
\ee
and similarly,
\be
\sum_{i=1}^{4} \vec{Q}_i^2=\sum_{i=1}^{5}\vec{p}_i^2,
\ee
and
\be
5\sum_{i=1}^{4} \vec{R}_i^2=\sum_{i,j,~i<j}(\vec{r}_i-\vec{r}_j)^2.
\ee

Apparently, if the harmonic oscillator potential between particles is taken into account, the five-particle problem can be solved analytically when the masses of particles are the same as each other.
Finally, the ground state energy of the five-particle system is
\be
\label{202501041348}
E^{(5)}=4 E^{(2)}\left(\mu \rightarrow m, \kappa \rightarrow 5\kappa \right)= 4\cdot \frac{3}{2} \hbar \sqrt{\frac{5\kappa}{m}}
=6\sqrt{5} \hbar \sqrt{\frac{\kappa}{m}}.
\ee
Moreover, it can be written as the summation of five oscillator energies, i.e.,
\be
E^{(5)}=5E^{(2)}\left(\mu \right),
\ee
with the reduced mass $\mu=\frac{5}{16}m$.

Just as the three-particle and four-particle cases, the root-mean-square radius of the ground state of the five-particle system is
\be
\label{eq:202501061802}
\sqrt{\langle \vec{r}^2 \rangle}=\sqrt{\frac{4}{5} \langle \vec{R}_1^2 \rangle}.
\ee
In the five-particle system, Eq.~(\ref{eq:01061514}) is still correct but $\alpha^2=\frac{\sqrt{5\kappa m}}{\hbar}$, Therefore,
\be
\langle \vec{R}_1^2 \rangle=\frac{9\hbar^2}{mE^{(5)}}.
\ee

\subsection{The $N$-particle system with a harmonic oscillator potential}
\label{sect:Nbody}

It is no doubt that this method can be extended to the system consisting of any particles with the same mass. In the $N-$ particle system, the ground state energy can be written as
\be
E^{(N)}=N \cdot E^{(2)}\left(\mu \right),
\ee
with $\mu=\frac{Nm}{(N-1)^2}$.

When the particle number increases, $N \rightarrow \infty$, the average energy per particle becomes divergent, $E^{(N)}/N \sim \sqrt{N}$, which implies this method can only be used to solve few-particle problems.

\section{Results}
\label{sect:results}

\subsection{The ground state energies of ${}^{12}C$, ${}^{16}O$ and ${}^{20}Ne$}
\label{sect:energy}

The Hoyle state of the ${}^{12}C$ nucleus can be treated as a system consisted of three alpha particles, or three Helium nuclei. The three alpha particles in the ${}^{12}C$ nucleus have the same mass and spin zero, and the $\alpha$ condensation is generated when all $\alpha$ particles lie in the lowest state.
If there exist the $\alpha$ condensate states of the ${}^{16}O$ and ${}^{20}Ne$ nuclei, they must be composed of four or five alpha particles respectively.
Therefore, it is reasonable to study the structure of the ${}^{12}C$, ${}^{16}O$ and ${}^{20}Ne$ nuclei with the method evaluated in Section~\ref{Framework}.

According to Eqs.~(\ref{eq:202412041008}),~(\ref{202501041347}) and (\ref{202501041348}), the ratio of the ground state energies of three-body, four-body and five-body systems can be written as
\be
\label{eq:202501041659}
E^{(3)}:E^{(4)}:E^{(5)}= 3\sqrt{3}  : 9 : 6\sqrt{5}\approx 1.732:3:4.472.
\ee
Actually, the Hoyle state of ${}^{12}C$ corresponds to the second $J^P=0^+$ state of ${}^{12}C$, which is 0.38MeV above the $3\alpha$ threshold. It implies the three-body energy of ${}^{12}C$ takes the value of
$E^{(3)}({}^{12}C)=0.38$MeV.
Analogy to the Hoyle state of ${}^{12}C$, the sixth $J^P=0^+$ state of ${}^{16}O$ is assumed to be a Hoyle-like state, which is composed of four $\alpha$ particles.
Since the sixth $J^P=0^+$ state of ${}^{16}O$ is 15.1MeV higher than the ground state of ${}^{16}O$ experimentally\cite{Schuck2008}, while the ground state of ${}^{16}O$ is 14.44MeV lower than the $4\alpha$ threshold\cite{Schuck2001}, the sixth $J^P=0^+$ state of ${}^{16}O$ composed of four $\alpha$ particles is only 0.66MeV above the $4\alpha$ threshold, i.e.,
$E^{(4)}({}^{16}O)=0.66MeV$.
Moreover, in Ref.~\cite{ZhouBo}, it is also asserted that the sixth $J^P=0^+$ state of ${}^{16}O$ might be an
$\alpha$ condensate state whose energy is less than 1MeV above the $4\alpha$ threshold. Therefore,
\be
E^{(3)}({}^{12}C):E^{(4)}({}^{16}O)=0.38:0.66\approx 1.727:3,
\ee
which is equal to the corresponding ratio of the three-body and four-body ground state energies in Eq.~(\ref{eq:202501041659}) approximately. At this point, the model derived in Section~\ref{Framework} can be used to study the structure of $4n$ nuclei, which are composed of $\alpha$ particles directly.

However, according to Eq.~(\ref{eq:202501041659}), the $\alpha$ condensate state of ${}^{20}Ne$ is about 1MeV above the $5\alpha$ threshold if the energies of $\alpha$ condensate states of ${}^{12}C$ and ${}^{16}O$ are correct.
Undoubtedly, it is far less than the experimental and theoretical energies of the $\alpha$ condensate state of ${}^{20}Ne$, which are 3MeV above the $5\alpha$ threshold at 19.2MeV\cite{ZhouBo,ZhouBo35,ZhouBo36}.

\subsection{The root-mean-square radii of ${}^{12}C$, ${}^{16}O$ and ${}^{20}Ne$}
\label{sect:radius}

The root-mean-square radii of ${}^{12}C$ and ${}^{16}O$ can be calculated according to Eqs.~(\ref{eq:202501061210}) and (\ref{eq:202501061458}), respectively. With the experimental ground state energies of ${}^{12}C$ and ${}^{16}O$, i.e., $E^{(3)}({}^{12}C)=0.38MeV$ and $E^{(4)}({}^{16}O)=0.66$, the root-mean-square radii of ${}^{12}C$ and ${}^{16}O$ equal to 9.08fm and 8.95fm, respectively, which are far larger than the corresponding value of ${}^{16}O$ in Ref.~\cite{Schuck2008}.
However, if the ground state energy of ${}^{16}O$ takes the value of 2MeV approximately, just as the result calculated in Ref.~\cite{Schuck2008}, the root-mean-square radius of ${}^{16}O$ is 5.14fm, which is close to the radius obtained in Ref.~\cite{Schuck2008}.
Apparently, the root-mean-square radius of the $\alpha$ condensate nucleus increases when it lies in a loosely bound state.
Furthermore, if the ${}^{12}C$ and ${}^{16}O$ nuclei lie in the $\alpha$ condensate states, the $\alpha$ particles wound distribute  as an equilateral triangle in ${}^{12}C$ and a regular tetrahedron in ${}^{16}O$, respectively.

As for the nucleus ${}^{20}Ne$, if the $\alpha$ particle are all located in the lowest energy state, the root-mean-square radius wound be 8.76fm with $E^{(5)}=0.98$MeV. When the ground state energy of ${}^{20}Ne$ takes the experimental value of 3MeV,  as given in Refs.~\cite{ZhouBo}, the root-mean-square radius reduces to 5.01fm, which is also displayed in Table~\ref{table:data}.

The experimental data on ${}^{12}C$, ${}^{16}O$ and ${}^{20}Ne$ are listed in Table~\ref{table:data}, respectively. The data on ${}^{12}C$ are taken from Ref.~\cite{Schuck2001}, and the data on ${}^{16}O$ come from Ref.~\cite{Schuck2008-34}, from Ref.~\cite{Schuck2008-35} for the $J=0_4^+$ state, from Ref.~\cite{ZhouBo35} for the $J=0^{+}_6$ state, respectively. As for those of ${}^{20}Ne$, most of them are extracted in Ref.~\cite{ZhouBo35}, while the energy $E=22.5$MeV
is from Ref.~\cite{ZhouBo36}, which is labeled with a dagger symbol in Table~\ref{table:data}.
In addition, the calculated energies $E^{(n)}=E-E_{n \alpha}^{th}$ and root-mean-square radii ${\sqrt{\langle r^2 \rangle}}$ of the $\alpha$ condensate states of these nuclei and the corresponding values in the THSR wave function are also included in Table~\ref{table:data}.

\section{Summary}
\label{sect:summary}

In this work, the three-body, four-body and five-body problems are solved analytically with a harmonic oscillator potential of particles by using an orthogonal transformation of the particle coordinates, and then the Hoyle state of ${}^{12}C$, the Hoyle-like  states of ${}^{16}O$ and ${}^{20}Ne$ are studied with this method. The ratio of the ground state energies of ${}^{12}C$ and ${}^{16}O$ is calculated, and the result agrees on that of the corresponding experimental values. Therefore, it implies that  the ground state energies of ${}^{12}C$ and ${}^{16}O$ can be obtained with the same harmonic oscillator potential of $\alpha$ particles. In other words, the parameters in the harmonic oscillator potential of $\alpha$ particles take the same value as each other whether it is determined with the experimental energy of ${}^{12}C$ or ${}^{16}O$. Nevertheless, it is not fit for ${}^{20}Ne$, the calculated ground state energy of ${}^{20}Ne$ is far less than the experimental value of 3MeV.
In addition, the root-mean-square radius of the $\alpha$ condensate state of ${}^{12}C$, ${}^{16}O$ and ${}^{20}Ne$ are calculated, all values are around 9fm, which is larger than the result of the THSR wave function. If a energy of 2MeV for ${}^{16}O$ or 3MeV for ${}^{20}Ne$ is adopted respectively, just as done in Refs.~\cite{Schuck2008} and~\cite{ZhouBo},
the root-mean-square radius would be around 5m, which is consistent with the result of the THSR wave function.

\begin{table}[htbp]
\begin{tabular}{c|c|ccc|cc|cc}
\hline \hline
                     &  State    & Experimental data &  &   &Our results &  &THSR results &    \\
\hline
                     &      & $E$ & $E-E_{n\alpha}^{th}$  &${\sqrt{\langle r^2 \rangle}}$  &$E-E_{n\alpha}^{th}$ & ${\sqrt{\langle r^2 \rangle}}$ &$E-E_{n\alpha}^{th}$ & ${\sqrt{\langle r^2 \rangle}}$   \\
                     &      & (MeV) & (MeV) & (fm)  &(MeV) & (fm) &(MeV) & (fm)   \\
\hline
 ${}^{12}C$           & $0_1^+$ & 0.00  & -7.27& 2.65 &      &    &-3.4& 2.97\\
                      & $0_2^+$ & 7.65  & 0.38 &      & 0.38 &9.08&0.5& 4.29\\
           & $E_{3\alpha}^{th}$ & 7.27  &      &      &      &    && \\
\hline
 ${}^{16}O$           & $0_1^+$   & 0.00   & -14.44 & 2.71  &  &&$\sim-14.4$&$\sim2.7$ \\
                      & $0_2^+$   & 6.05  & -8.39 &         &   &&$\sim-8.0$&$\sim3.0$\\
                      & $0_3^+$   &12.1   & -2.34 &        &   &&$\sim-4.0$&$\sim3.0$\\
                      & $0_4^+$   &13.6  & -0.84&         &   &&$\sim-2.0$&$\sim4.0$\\
                      & $0_5^+$   &14.0  & -0.44 &          &   &&$\sim0.0$&$\sim3.1$\\
                      & $0_6^+$   & $15.1$  & 0.66 &      & 0.66  &8.95&$\sim2.0$&$\sim5.0$\\
           &  $E_{4\alpha}^{th}$  & 14.44    &  &           &   &&&\\
\hline
 ${}^{20}Ne$          &           & 21.2     & 2.0 &   & 0.98 &8.76&&\\
                      &           & 21.8     & 2.6 &   & $3.0$  &5.01&$\sim3.0$&\\
                      &           & 23.6     & 4.4 &   &      &    &&\\
                      &           & $22.5^\dagger$     & 3.3 &   &      &    &&\\
           &  $E_{5\alpha}^{th}$  & 19.2     &     &   &      &    &&\\
\hline \hline
\end{tabular}
\caption{The experimental data, the calculated energies $E^{(n)}=E-E_{n \alpha}^{th}$ and root-mean-square radii ${\sqrt{\langle r^2 \rangle}}$ of atomic nuclei ${}^{12}C$, ${}^{16}O$ and ${}^{20}Ne$ and the corresponding results of the THSR wave function, where $E_{n \alpha}^{th}$ is the $n \alpha$ threshold. The detailed discussion can be found in the text.} \label{table:data}
\end{table}

\begin{acknowledgments}
Bao-Xi Sun would like to thank Jean-Marc Richard for useful discussions. 
\end{acknowledgments}

\newpage


\begin{thebibliography}{99}

\bibitem{Hoyle1954}
F. Hoyle, Astrophys. J. Suppl. {\bf 1}, 121 (1954).

\bibitem{Schuck2008-4}
F. Dalfovo, S. Giorgini, L. P. Pitaevskii and S. Stringari, Rev. Mod. Phys. {\bf 71}, 463 (1999).

\bibitem{Schuck2001}
A. Tohsaki, H. Horiuchi, P. Schuck and G. Ropke, Phys. Rev. Lett. {\bf 87}, 192501 (2001).

\bibitem{Schuck2008}
Y.Funaki, T. Yamada, H. Horiuchi, G. Ropke, P. Schuck and A. Tohsaki, Phys. Rev. Lett. {\bf 101}, 082502 (2008).

\bibitem{Schuck2008-34}
F. Ajzenberg-Selove, Nucl. Phys. A {\bf 460}, 1 (1986).

\bibitem{Schuck2008-35}
T. Wakasa et al., Phys. Lett. B {\bf 653}, 173 (2007).

\bibitem{ZhouBo35}
S. Adachi et al., Phys. Lett. B {\bf 819}, 136411 (2021).

\bibitem{ZhouBo36}
J. A. Swartz et al., Phys. Rev. C {\bf 91}, 034317 (2015).

\bibitem{ZhouBo}
B. Zhou, Y.Funaki, H. Horiuchi, Y. G. Ma, G. Ropke, P. Schuck, A. Tohsaki and T. Yamada, Nature Commun. {\bf 14}, 8206 (2023).

\bibitem{Richard90-1}
J.-L. Basdevant, J.-M. Richard and A. Martin, Nucl. Phys. B {\bf 343}, 60 (1990).

\bibitem{Richard90-2}
J.-L. Basdevant, J.-M. Richard and A. Martin, Nucl. Phys. B {\bf 343}, 69 (1990).

\bibitem{Richard93}
J.-L. Basdevant, J.-M. Richard, A. Martin and T. T. Wu, Nucl. Phys. B {\bf 393}, 111 (1993).

\bibitem{Basdevant96}
J.-L. Basdevant and A. Martin, J. Math. Phys. {\bf 37}, 5916 (1996).

\bibitem{Book}
J.-L. Basdevant and J. Dalibard, {\sl The quantum mechanics solver: How to apply quantum theory to modern physics}(3rd edition), Springer Nature Switzerland, 2019.

\end{thebibliography}
\end{document}